\documentclass[11pt]{article}

\usepackage{amsthm}
\usepackage{graphicx}
\usepackage[utf8]{inputenc}
\usepackage{subfig} 
\usepackage{url}

\usepackage[english]{babel}
\usepackage{amsmath}
\usepackage{amsfonts}
\usepackage{amssymb}
\usepackage{xcolor}
\usepackage{fourier}
\usepackage{hyperref}
\usepackage[capitalise]{cleveref}
\usepackage{lscape}
\usepackage{rotating}

\newcommand{\E}{\mathbb{E}}

\newcommand{\din}{k}

\newcommand{\pa}{\pi^{(1)}_t}
\newcommand{\ra}{\pi^{(2)}_t}

\numberwithin{equation}{section}

\usepackage{dcolumn,amsthm}

\renewenvironment{proof}[1][Proof]{\noindent\textit{#1. } }{\hfill$\square$}

\newtheoremstyle{theorem}{6pt}{6pt}{\rm}{}{\sffamily}{ }{ }{}
\theoremstyle{theorem}
\newtheorem{theorem}{\sc Theorem}[section]

\newtheoremstyle{lemma}{6pt}{6pt}{\rm}{}{\sffamily}{ }{ }{}
\theoremstyle{lemma}

\newtheoremstyle{example}{6pt}{6pt}{\rm}{}{\sffamily}{ }{ }{}
\theoremstyle{example}

\newtheoremstyle{corollary}{6pt}{6pt}{\rm}{}{\sffamily}{ }{ }{}
\theoremstyle{corollary}
\newtheorem{corollary}{\sc Corollary}[section]

\newtheoremstyle{definition}{6pt}{6pt}{\rm}{}{\sffamily}{ }{ }{}
\theoremstyle{definition}
\newtheorem{definition}[theorem]{\sc Definition}

\newtheoremstyle{remark}{6pt}{6pt}{\rm}{}{\sffamily}{ }{ }{}
\theoremstyle{remark}

\theoremstyle{plain}

\graphicspath{ {./images/} }

\author{\sc{Jan Medina-López} and \sc{Jorge Finke}\\[5pt]
Pontificia Universidad Javeriana, Cali, Colombia\\[3pt]
Corresponding author: janmedina@javerianacali.edu.co
}

\title{Characterizing the head of the degree distributions of growing networks}
 
\begin{document}
\maketitle  

\begin{abstract}
The analysis in this paper helps to explain the formation of growing networks with degree distributions that follow extended exponential or power-law tails. We present a generic model in which edge dynamics are driven by a continuous attachment of new nodes and a mixed attachment mechanism that triggers random or preferential attachment. Furthermore, reciprocal edges to newly added nodes are established according to a response mechanism. The proposed framework extends previous mixed attachment models by allowing the number of new edges to vary according to various discrete probability distributions, including Poisson, Binomial, Zeta, and Log-Series. We derive analytical expressions for the limit in-degree distribution that results from the mixed attachment and response mechanisms. Moreover, we describe the evolution of the dynamics of the cumulative in-degree distribution. Simulation results illustrate how the number of new edges and the process of reciprocity significantly impact the head of the degree distribution.
\end{abstract}
\section{Introduction}\label{section1}
As new nodes become part of a network, multiple mechanisms contribute to the evolution of the degree of each node~\cite{Ghoshal2013}. The degree distribution $P(k)$ that results from such mechanisms often ranges orders of magnitude in values of $k$~\cite{Ghoshal2013,Barabasi1999}. In particular, the work in~\cite{Barabasi1999} characterizes the probability of encountering nodes with degree $k$ in networks with power-law distributions of the form $P(k)\sim k^{-\gamma}$ for $\gamma = 3$. The authors explain power-law distributions as an outcome of two mechanisms: \emph{constant growth}, which assumes that every time step a new node is added and establishes a fixed number of edges (i.e., the number of new connections follows a constant distribution), and \emph{attachment}, under which new nodes create new edges to existing target nodes (i.e., in the case of \emph{linear preferential attachment}, in direct proportion to the degree of the target nodes). For scenarios in which new nodes join the network according to the constant growth mechanism but without preference for nodes with a high degree (i.e., the attachment mechanism is random), the resulting distribution follows an exponential form, that is, $P(k)\sim \exp\left(-\lambda k\right)$ for $\lambda > 0$. Many empirical networks, however, do not exhibit such pure power-law or exponential relationships~\cite{Broido2018}.\\

The work in~\cite{Redner1998} shows that the degree distributions of scientific citations, for example, follows extended power-law distributions. Extended power-laws are characterized by the more general form $P(k)=\left(k+k_{\mathrm{sat}}\right)^{\gamma}$. While the tail of the degree distribution reduces to the pure power-law form (i.e., for degrees that are larger than the threshold $k_{\mathrm{sat}}$), the head degenerates into the pure exponential form (i.e., for degrees smaller than $k_{\mathrm{sat}}$).\\

Past studies illustrate how extended power-law distributions may result from the combination of a constant growth mechanism and a mixed attachment mechanism (which itself involves random and preferential attachment)~\cite{Shao2006,Medina2019}. In particular, the work in~\cite{Shao2006} shows that the relative contribution of random and preferential attachment defines the power-law exponent~$\gamma$. Meanwhile, the growth mechanism only influences the threshold $k_{\mathrm{sat}}$. More recently, related studies analyze the effect of response mechanisms on $k_{\mathrm{sat}}$ for directed networks~\cite{Fernandez2015,Valverde2007}. The work in~\cite{Fernandez2015}, for example, presents a model that considers two \emph{response mechanisms}. The first mechanism allows target nodes to respond to node attachment by establishing reciprocal edges. The second mechanism allows randomly selected nodes to establish edges to the newly added nodes. The analysis in~\cite{Fernandez2015} shows how each response mechanism ultimately impacts $k_{\mathrm{sat}}$. The authors also quantify the effects of the response mechanisms on the exponent of the exponential or power-law function that governs the tail of the in-degree distribution.\\ 

Similarly, the work in~\cite{Valverde2007} introduces a model that combines mechanisms of preferential attachment with node aging to generate extended power-laws. The authors show that node aging impacts both distribution parameters $k_{\mathrm{sat}}$ and $\gamma$. All the above models~\cite{Shao2006,Fernandez2015,Valverde2007} produce particular extended power-laws, which are characterized by the exponential behavior for degrees less than $k_{\mathrm{sat}}$. However, they fall short in resembling extended power-laws that follow other functional forms for the head of the distribution.\\

This paper extends the work in~\cite{Shao2006,Fernandez2015} by considering scenarios in which the number of new edges varies according to a random variable that obeys a particular probability distribution (with a fixed mean). The proposed model explains the formation of networks that follow piecewise exponential or power-law degree distributions. That is, for $k\geq k_{\mathrm{sat}}$, the resulting distribution follows an exponential or power-law distribution. For $k< k_{\mathrm{sat}}$, the behavior of $P(k)$ is governed by the probability distribution associated with the number of new edges established during attachment. As in~\cite{Shao2006}, target nodes are selected according to mixed attachment. As in~\cite{Fernandez2015}, a reciprocal response mechanism allows existing nodes to establish edges to new nodes. Unlike~\cite{Fernandez2015}, varying the number of new edges established by the attachment mechanism (i.e., non-constant edge growth) enables the model to resemble the head of various classes of empirical distributions, including networks of references between entries in dictionaries.\\ 

The contributions of this paper are the following. First, we untangle the effects of non-constant edge growth, mixed attachment, and response mechanisms. In particular, we extend the work in~\cite{Shao2006,Fernandez2015} by allowing the number of new edges to vary as networks grow by the continuous addition of nodes. Second, we characterize the probability that the new node has a particular in-degree (\cref{theorem3.1}). Furthermore, we characterize the in-degree distribution as a function of the parameters of each mechanism (Theorem~\ref{theorem3.2}). Finally, we show how the proposed model resembles two empirical networks, namely, the network of references between entries in the Free On-line Dictionary of Computing (FOLDOC) and the network of interactions of questions and answers on Stack Overflow.\\

The remainder of the paper is organized as follows. \Cref{section2} introduces the mixed attachment model with reciprocity. \Cref{section3} characterizes the in-degree distribution of networks that results from the proposed mechanisms. \cref{section4} evaluates the extent to which the model resembles the behavior of the two empirical networks. It also characterizes the impact of the number of new edges and reciprocity over the head and tail of the in-degree distribution. Finally, \Cref{section5} draws some concluding remarks and future research directions.%section1
\section{Network Model}\label{section2}
Let $I=\{0,1,2,\ldots\}$.  Consider a sequence of graphs $\left\{G_t\right\}_{t\in I}$ where each graph $G_t=(V_t,E_t)$ represents a directed network with a set of nodes $V_t$ and a set of edges $E_t \subseteq V_t \times V_t$ at time index~$t$. A pair $(u,v)\in E_t$ represents a directed edge from \textit{source} node $u$ to \textit{target} node $v$. Let the expression $\din_t(u)$ denote the \textit{in-degree} of node $u\in V_t$. The expressions $n_t=|V_t|$ and $e_t=|E_t|$ denote the \textit{number of nodes} and the \textit{number of edges} in $G_t$.\\ 

Let $0\leq p\leq 1$ denote the extent to which preferential attachment dominates the attachment process. Moreover, let $A$ and $R$ denote random variables that characterize the number of new edges established due to mixed attachment and reciprocity mechanisms, respectively. The expressions $a_n$ and $\E[A]$ denote the $n\mathrm{th}$ realization and the expected value of $A$. Finally, let $M$ be a random variable that describes the total number of new edges (established by the new and existing nodes).

\begin{definition}\label{definition3.1}
The network model starts from a seed network $G_0$ with $v_0$ nodes and $e_0$ edges. At each time $t>0$, the network evolves as follows: 
    \begin{enumerate}
        \item[M1] \textit{Growth}. A new node is added to $V_{t-1}$. 
        
        \item[M2] \textit{Mixed Attachment}. For a fixed $p$, the new node selects and attaches to $a_t$ different nodes in $V_{t-1}$ according to the probability
        \begin{equation}
          \pi_t(v\mid p)= p\pa (v) + (1-p)\ra (v), \label{eq2.1}
        \end{equation}
        where $0\leq p\leq 1$ and $\pa (v)$ represents the probability of being chosen as a target node due to preferential attachment, i.e.,
          \begin{equation}
            \pa(v)=\frac{\din_{t-1}(v)}{\sum\limits_{w \in V_{t-1}}\din_{t-1}(w)}, \label{eq2.2}
          \end{equation}
        and $\ra(v)$ represents the probability of being chosen as a target node due to random attachment, i.e.,
          \begin{equation}
            \ra(v)=\frac{1}{n_{t-1}}. \label{eq2.3}
          \end{equation}
        \item[M3] \textit{Reciprocity}. Finally, each of the $a_i$ selected nodes establishes an edge to the new node with probability $0< q\leq 1$.
    \end{enumerate} 
\end{definition}

Mechanisms M1-M3 are iterated until a desired number of nodes has been added to the network. To ensure that mechanisms M1-M3 are well-defined, the seed network must have at least $a_1+1$ nodes. This assumption guarantees that the node added at $t=1$ can establish up to $a_1$ new edges.\\

Note that mechanism M3 follows a finite Bernoulli process, so each edge established due to mechanism M3 is characterized as a Bernoulli trial with parameter~$q$. Since mechanism M1 creates $a_t$ new edges at each time $t$, the expected number of Bernoulli trials is determined by $\E[A]$. Hence, there are $\E[A]$ statistically independent Bernoulli trials, each with a probability of success $q$. Therefore, the variable $R$ is binomially distributed with parameters $\E[A]$ and $q$, and expected value $\E[R] = q \E[A]$. Furthermore, since $A$ and $R$ are independent, the expected number of new edges is $m = \E[M] = (1 + q)\E[A]$.%section2
\section{Asymptotic Behavior of the In-Degree Distribution}\label{section3}
This section characterizes the in-degree distribution of a network generated by the model presented in \cref{section2}, starting from a given seed network $G_0$. It also shows that the dynamics of the in-degree distribution converges to a stationary distribution. We conclude this section by presenting simulations that validate our analytical results.\\

Let $K_t$ be a random variable that characterizes the in-degree of a node selected uniformly at random from $G_t$. Moreover, let $P_t(k)=\mathbb{P}(K_t=k)$ denote the probability that a realization of $K_t$ equals $k$. The expression $F_t(k) = \mathbb{P}(K_t > k) = 1 - \sum_{i\leq k} P_t(i)$ denotes the complementary cumulative distribution function. Finally, let $\mathbb{B}(k)$ denote the probability that a new node has in-degree $k$.\\

Note that, both the support of the probability distribution of $A$ and mechanism M3 affect the minimum degree of all nodes across the network. When mechanism M3 is triggered for each new attachment ($q=1$), the minimum degree present in the network depends solely on the support of the distribution of $A$. In contrast, if the reciprocal response varies ($0<q<1$), the minimum in-degree is zero. Therefore, the probability that the new node has in-degree equal to $k$ is given by
\begin{enumerate}
    \item If~$0<q<1$,
    \begin{equation}
        \mathbb{B}(k)= 
        \sum_{i=k}^{\infty}\binom{i}{k}q^k(1-q)^i~\mathbb{P}(A=i).
    \label{eq3.1}
    \end{equation}
    \item If $q=1$,
    \begin{equation}
        \mathbb{B}(k)=\mathbb{P}(A=k).
    \label{eq3.2}
    \end{equation}
\end{enumerate}
For the case where $0<q<1$, the probability distribution $\mathbb{B}$ is a series that depends on the distribution of $A$. The following theorem presents conditions under which \cref{eq3.1} converges. It also provides a closed-form for the convergent series for different distributions of~$A$.
\begin{theorem}\label{theorem3.1}
The probability that the new node has in-degree $k$ (\cref{eq3.1}) converges to the following expressions.
\begin{enumerate}
\item If $A\sim\mathsf{Poisson}(r)$ with $r>0$, then $\mathbb{B}(k)=\frac{(q\theta)^k}{k!}e^{q\theta}$.

\item If $A\sim\mathsf{Binomial}(r,~\theta)$ and $0\leq\frac{\theta}{1-\theta}(1-q)<1$, then $\mathbb{B}(k)=\binom{r}{k}(q\theta)^k(1-q\theta)^{r-k}$.
\item If $A\sim\mathsf{Geometric}(\theta)$, then $\mathbb{B}(k)=\frac{\theta}{q+\theta(1-q)}\left(\frac{q(1-\theta)}{q+\theta(1-q)}\right)^k$.
\item If $A\sim\mathsf{NegativeBinomial}(r,~\theta)$, then $
\mathbb{B}(k)=\binom{k+r-1}{r-1}\left(\frac{q(1-\theta)}{q+\theta(1-q)}\right)^k\left(\frac{\theta}{q+\theta(1-q)}\right)^r$.
\item If $A\sim\mathsf{Zeta}(r)$ with $r\geq 1$, then
\begin{equation*}
\mathbb{B}(k)=\left\{ \begin{array}{lll}
           \frac{1}{\zeta(r+1)}\mathrm{Li}_{r+1}(1-q)&, &k=0 \\
             \\ 
             \frac{1}{\zeta(r+1)k!}\left(\frac{q}{1-q}\right)^k\sum_{i=1}^{k}e_{i-1}(1,\ldots,k)\mathrm{Li}_{r+i-k}(1-q) &, &k\geq 1
             \end{array}
             \right.
\end{equation*}
where $\mathrm{Li}_{r}(x)$ is the poly-logarithm function and $e_i(1,\ldots,k)$ denotes the elementary symmetric polynomials in $k$ variables.

%need to talk about this sub sub indices

\item If $A\sim\log(\theta)$, then 
\begin{equation*}
\mathbb{B}(k)=\left\{ \begin{array}{lll}
           \frac{\ln\left(1+(q-1)\theta\right)}{\ln(1-\theta)}&, &k=0 \\
           \\ 
            \frac{-1}{k\ln(1-\theta)}\left(\frac{q\theta}{1-\theta(1-q)}\right)^k &, &k\geq 1
            \end{array}
             \right.
\end{equation*}

\end{enumerate}
\end{theorem}

\cref{theorem3.1} shows the relationship between the in-degree of a new node and the distribution of the number of new edges. The following theorem guarantees the convergence of the in-degree probability distribution as the network grows.

\begin{theorem}\label{theorem3.2}
If $0\leq p\leq 1$ and $0 < \E[A] < \infty$, then $\lim\limits_{t\rightarrow\infty}P_t(k)$ exists for all $k\geq 0$.
\end{theorem}
\begin{proof}
For the sake of simplicity we assume that the minimum in-degree of all nodes across the network is zero. Following a similar argument as in~\cite{Dorogovtsev2000}, we want to determine a recursive expression for $P_t(k)$ for $k \geq 0$.\\

According to \cref{eq2.1,eq2.2,eq2.3}, for a fixed $p$, the probability that the new node connects to a target node $v$ with in-degree $k$ is
\begin{align}
    \E[A]\pi_t(k)=p\frac{k\E[A]}{e_{t-1}} + (1-p)\frac{\E[A]}{n_{t-1}}.\label{eq3.3}
\end{align}
According to \cref{eq3.3} for $k > 0$, the expected number of nodes with in-degree $k$ is given by
\begin{align}
    \E[A]\pi_t(k)n_{t-1}P_{t-1}(k) = \left(\frac{p k n_{t-1}\E[A]}{e_{t-1}}+(1-p)\E[A]\right)P_{t-1}(k).\label{eq3.4}
\end{align}
Using \cref{eq3.4} the expected number of nodes with in-degree zero is given by
\begin{equation}
n_{t}P_{t}(0) = n_{t-1}P_{t-1}(0)-\pi_t(0)\E[A]n_{t-1}P_{t-1}(0)+\mathbb{B}(0).
\label{eq3.5}
\end{equation}
The first term in \cref{eq3.5} represents the expected number of nodes with no incoming edges at time $t-1$. The second term corresponds to the expected number of nodes without incoming edges that establish an edge with the new node at time $t$. Finally, $\mathbb{B}(0)$ accounts for the expected new nodes attaching to the network and having no incoming edges. Now, the expected number of nodes with in-degree $k>0$ is given by
  \begin{align}
    n_{t}P_{t}(k) =&~n_{t-1}P_{t-1}(k)-\pi_t\left(k\right)\E[A]n_{t-1}P_{t-1}(k)+\pi_t\left(k-1\right)\E[A]n_{t-1}P_{t-1}(k-1)+\mathbb{B}(k).
    \label{eq3.6}
  \end{align}
The first term in \cref{eq3.6} represents the expected number of nodes with in-degree $k$ at time $t-1$. The second term represents the expected number of nodes with in-degree $k$ selected at time $t-1$ by mechanism M2. The third term represents the expected number of nodes with in-degree $k-1$ that establish an edge with the new node. Finally, the fourth term represents the probability that the new node has in-degree $k$ (given by mechanism M3).\\

Proving the existence of the limit follows by induction over $k$.
\begin{itemize}
\item[] \textbf{Base Case}. When $k=0$, using \cref{eq3.5}, $P_t(0)$ can be expressed using the recurrence
  \begin{equation}
P_t(0)=\alpha_t P_{t-1}(0)+\beta_t,\label{eq3.7}
  \end{equation}
where $\alpha_t=1-n_t^{-1}\left(1+\pi_t(0)\E[A]n_{t-1}\right)$ and $\beta_t=n_t^{-1}\mathbb{B}(0)$. The expression in \cref{eq3.7} is a non-autonomous, first-order difference equation with initial condition $P_0(0)\in[0,1]$. By induction over $t$, it can be shown  that the solution to \cref{eq3.7} is given by
  \begin{equation}
P_t(0)=\prod\limits_{i=1}^{t}\alpha_iP_0(0)+\sum\limits_{i=1}^{t}\prod\limits_{j=i+1}^{t}\beta_i\alpha_j.\label{eq3.8}
  \end{equation}
Since $\sum_{t=1}^{\infty}\log\left(n_t^{-1}\left(1+\pi_t(0)\E[A]\right)\right)$ diverges as $t\rightarrow\infty$, the first term in \cref{eq3.8} diverges. Moreover, as $t\rightarrow\infty$, the second term in \cref{eq3.8} is a convergent series given by
\[\sum\limits_{i=1}^{\infty}\prod\limits_{j=i+1}^{\infty}\beta_j\alpha_i=\frac{\mathbb{B}(0)}{1 + \E[A](1-p)},\]
where $m = (1 + q)\E[A]$ denotes the expected number of new edges. Therefore,
\[\lim\limits_{t\rightarrow\infty}P_t(0)=\frac{\mathbb{B}(0)}{1 + \E[A](1-p)}.\]

\item[] \textbf{Inductive step.} Let $k>0$ and assume that $\lim\limits_{t\rightarrow\infty}P_t(k)$ exists. Using \cref{eq3.4,eq3.6} we have
\begin{equation}
\label{eq3.9}
\begin{split}
n_{t}P_{t}(k+1) =& \left(n_{t-1}-\frac{p (k+1) n_{t-1}\E[A]}{e_{t-1}}-(1-p)\E[A]\right)P_{t-1}(k+1)\\
&+\left(\frac{p k n_{t-1}\E[A]}{e_{t-1}}+(1-p)\E[A]\right)P_{t-1}(k)+\mathbb{B}(k+1).
\end{split}
\end{equation}
Using \cref{eq3.9} and \cite[Lemma 2]{Ruiz2018}, for a large enough $t$
\begin{align*}
\left(\frac{1+q+p (k+1)+m(1-p)}{1+q}\right)P_t(k+1) \sim \left(\frac{p k+m(1-p)}{1+q}\right)P_t(k)+\mathbb{B}(k+1).
\end{align*}
By our inductive hypothesis
\begin{equation*}
\lim_{t\rightarrow\infty}P_t(k+1)=\frac{p k+m(1-p)}{1+q+p (k+1)+m(1-p)}\lim_{t\rightarrow\infty}P_t(k)+\frac{(1+q)\mathbb{B}(k+1)}{1+q+p (k+1)+m(1-p)}.
\end{equation*}
\end{itemize}
\end{proof}

By defining $P(k)=\lim\limits_{t\rightarrow\infty}P_t(k)$, and applying \cref{theorem3.2} for a particular in-degree $k$, we can characterize the behavior of the in-degree distribution as 
\begin{equation}
P(k)= \left\{ \begin{array}{lll}
             \dfrac{\mathbb{B}(0)}{1 +\E[A](1-p)} &   ,  & k = 0 \\
             \\ 
             \dfrac{p (k-1)+m(1-p)}{1+q+p k+m(1-p)}P(k-1)+\dfrac{(1+q)\mathbb{B}(k)}{1+q+p k+m(1-p)}&  ,  & k> 0
             \end{array}
   \right.
   \label{eq3.13}
\end{equation}
It can be shown that the solution of \cref{eq3.13} is given by
\begin{enumerate}
    \item For $p=0$,
    \begin{equation}
        P(k)=\frac{1}{1 + \E[A]}\sum\limits_{j=0}^{k}\left(\frac{\E[A]}{1+\E[A]}\right)^{k-j}\mathbb{B}(j),\quad k\geq 0.
        \label{eq3.14}
    \end{equation}
    \item For $0<p<1$, 
    \begin{equation}
       P(k)=\frac{\Gamma\left(k+\frac{m(1-p)}{p}\right)}{\Gamma\left(k+\frac{1+q+p+m(1-p)}{p}\right)}\sum\limits_{j=0}^{k}\frac{(1+q)\Gamma\left(j+\frac{1+q+p+m(1-p)}{p}\right)}{\left((1 + q)(1+\E[A](1-p)) + pj\right)\Gamma\left(j+\frac{m(1-p)}{p}\right)}\mathbb{B}(j).
        \label{eq3.15}
    \end{equation}
    \item For $p=1$,
    \begin{equation}
       P(k)=\left\{ \begin{array}{lll}
             \mathbb{B}(0) & , & k = 0 \\
             \\             \frac{(1+q)\Gamma(k)}{\Gamma(k+q+2)}\sum\limits_{j=1}^{k}\frac{\Gamma(j+q+2)}{(j+q+1)\Gamma(j)}\mathbb{B}(j)& , & k> 0
             \end{array}
             \right.
        \label{eq3.16}
    \end{equation}
\end{enumerate}

Moreover, using \cref{eq3.14,eq3.15,eq3.16}, the following corollary characterizes the limit behavior of the complementary cumulative distribution function (CCDF) of the in-degree.

\begin{corollary}
Let $B(x,y)$ denotes the beta function. The CCDF of the in-degree converges and its asymptotic behavior satisfies
\begin{enumerate}
    \item For $p=0$,
    \begin{equation}
         F(k)=1-\frac{1}{1 + \E[A]}\sum\limits_{i=0}^{k-1}\sum\limits_{j=0}^{i}\mathbb{B}(j)\left(\frac{\E[A]}{1+\E[A]}\right)^{i-j}.
         \label{eq3.10}
    \end{equation}
    \item For $0<p<1$,
    \begin{equation}
        F(k)=1-\sum\limits_{i=0}^{k-1}\sum\limits_{j=0}^{i}\frac{\left(1+q\right)B\left(i+\frac{m(1-p)}{p},\frac{1+q+p}{p}\right)}{\left(1 + q + pi+m(1-p)\right)B\left(j+\frac{m(1-p)}{p},\frac{1+q+p}{p}\right)}\mathbb{B}(j).
        \label{eq3.11}
    \end{equation}
    \item For $p=1$,
    \begin{equation}
        1-\sum\limits_{i=0}^{k-1}\sum\limits_{j=1}^{k}\frac{(1+q)B\left(i,q+2\right)}{(j+q+1)B\left(j,q+2\right)}\mathbb{B}(j).
        \label{eq3.12}
    \end{equation}
\end{enumerate}
\end{corollary}

Next, the following theorem characterizes the dynamics of the complementary cumulative in-degree distribution.

\begin{theorem}\label{theorem3.3}
For $t\geq 1$ and $0<q<1$, the evolution of the complementary cumulative in-degree distribution satisfies
\begin{equation}
F_t(k)=\left\{
    \begin{array}{lcc}            \frac{\E[A](1-p)}{n_t}+\frac{n_{t-1}-\E[A](1-p)}{n_t}F_{t-1}(k)+\frac{1-\mathbb{B}(0)}{n_t} & , & k=0\\
             \\ 
             \frac{n_{t-1}\E[A]\pi_t(k)}{n_t}F_{t-1}(k-1)+\frac{n_{t-1}\left(1-\E[A]\pi_t(k)\right)}{n_t}F_{t-1}(k)+\frac{1-\sum_{i=0}^{k}\mathbb{B}(k)}{n_t} & , & k=0
             \end{array}
    \right.
\end{equation}
\end{theorem}
\begin{proof}
We know that mechanisms M1 and M3 increase $F_t(k)$ when new edges are established to nodes that have in-degree $k$. In particular, mechanism M1 increases $F_t(k)$ by
\begin{equation}
\E[A]\pi_t(k)P(k)=\E[A]\pi_t(k)\left(F_{t}(k-1)-F_{t}(k)\right) \label{eq4.18}
\end{equation}
Now, consider the effect of mechanism M3 when it establishes new edges from existing nodes to new nodes. It is clear that mechanism M3 only affects the in-degree of the new node. In particular, the probability that the new node has in-degree equal to $k$ is given by \cref{eq3.1}. Since $\mathbb{B}$ is a discrete probability distribution, the probability that a new node has in-degree greater than $k$ can be rewritten as
\begin{equation}
    \sum_{i=k+1}^{\infty} \mathbb{B}(i)=1-\sum_{i=0}^{k}\mathbb{B}(i)\label{eq4.19}
\end{equation}
Therefore, according to \cref{eq4.18,eq4.19} the expected number of nodes with in-degree $k=0$ is
\begin{equation*}
n_t F_t(0)=n_{t-1}F_{t-1}(0)+n_{t-1}\pi_t(0)\E[A]\left(1 - F_{t}(0)\right)+1-\mathbb{B}(0),
\end{equation*}
and the expected number of nodes with in-degree $k>0$ can be characterized as
\begin{equation*}
n_t F_t(k)=n_{t-1} F_{t-1}(k)+n_{t-1}\pi_t(k)\E[A]\left(F_{t-1}(k)-F_{t}(k+1)\right)+1-\sum_{i=0}^{k}\mathbb{B}(i).
\end{equation*}
\end{proof}

\begin{figure}[!hbtp]
    \centering
    \subfloat[][]{\includegraphics[width=200pt,height=120pt]{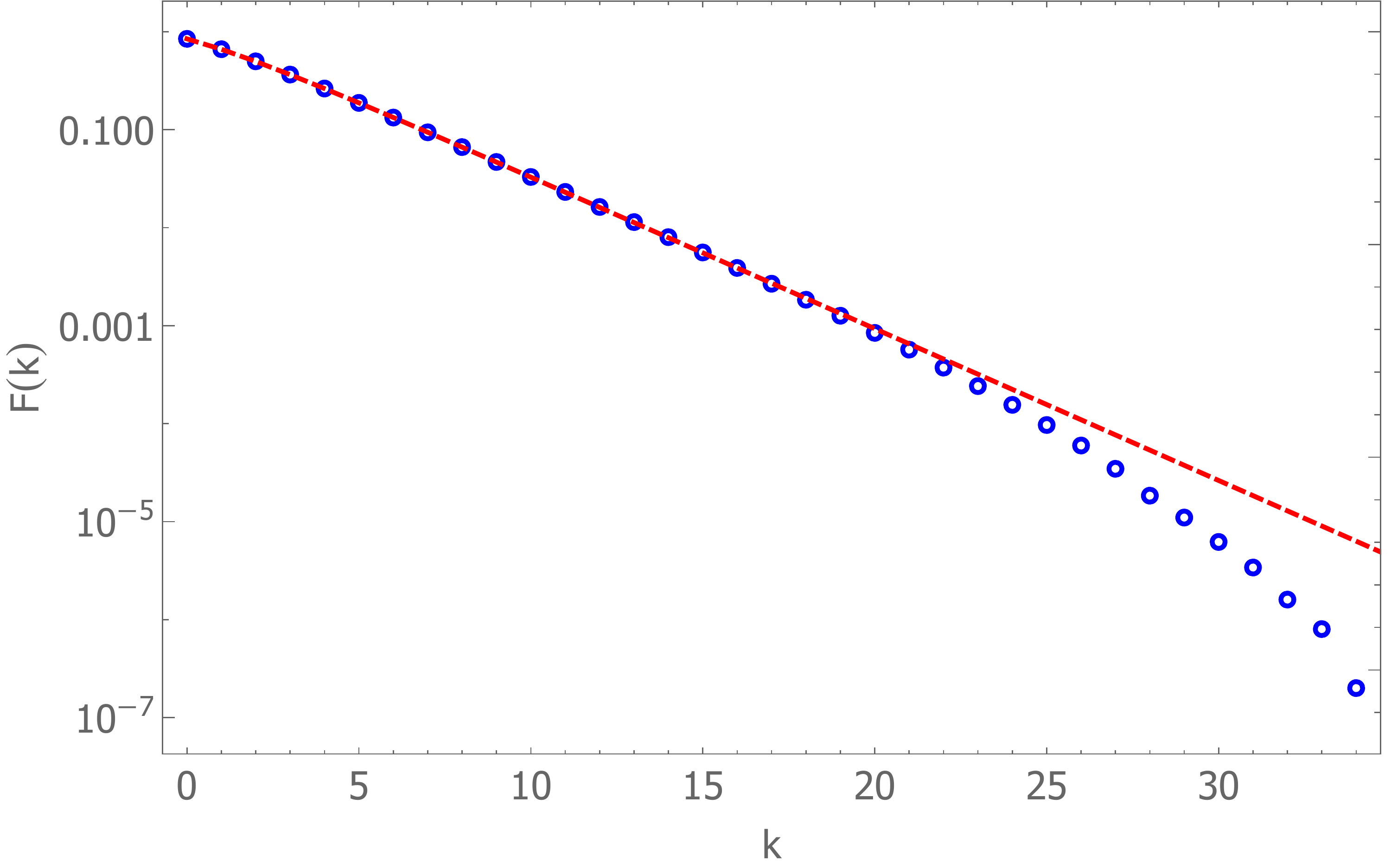}}~
    \subfloat[][]{\includegraphics[width=200pt,height=120pt]{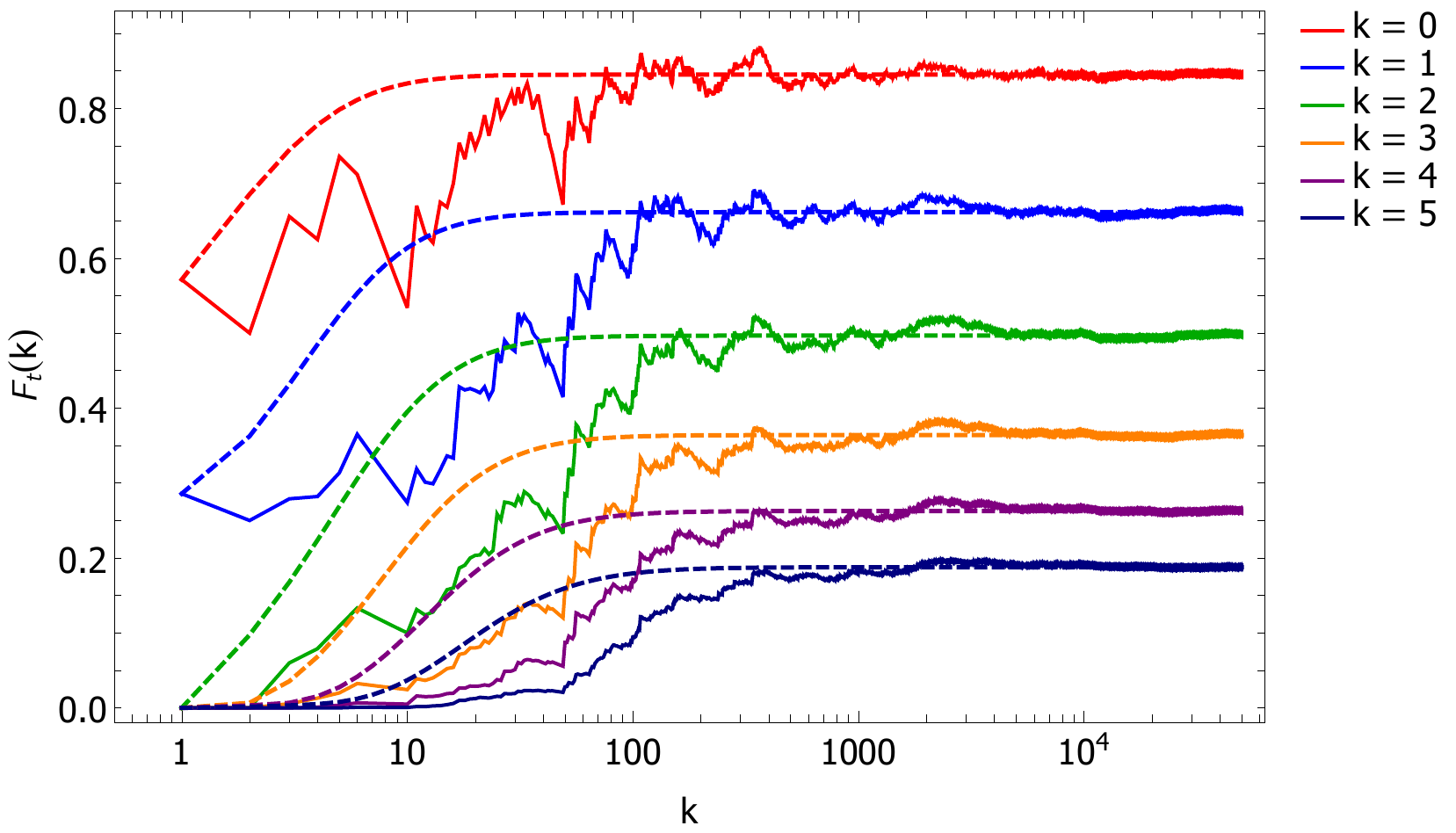}}
    
    \subfloat[][]{\includegraphics[width=200pt,height=120pt]{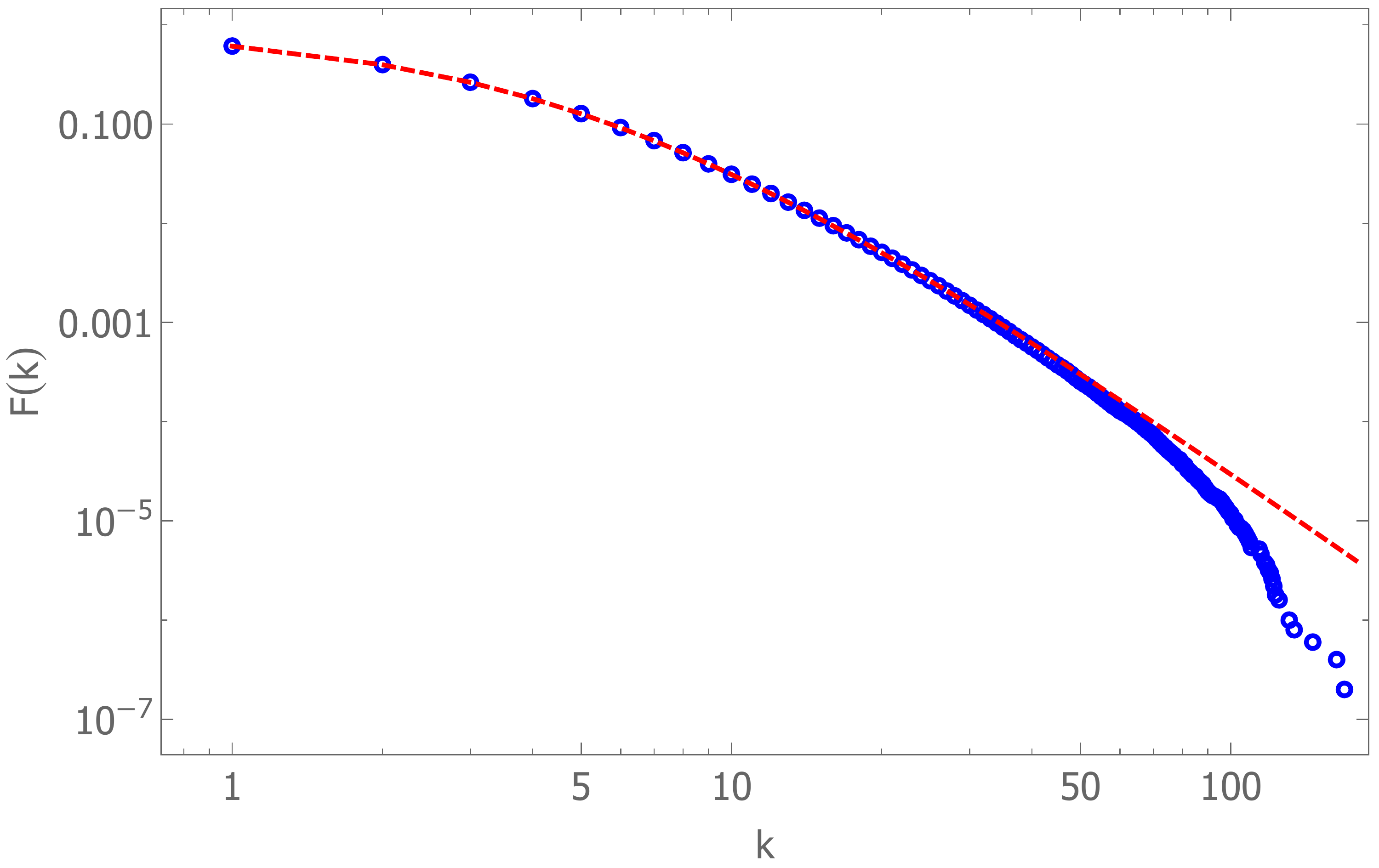}}~
    \subfloat[][]{\includegraphics[width=200pt,height=120pt]{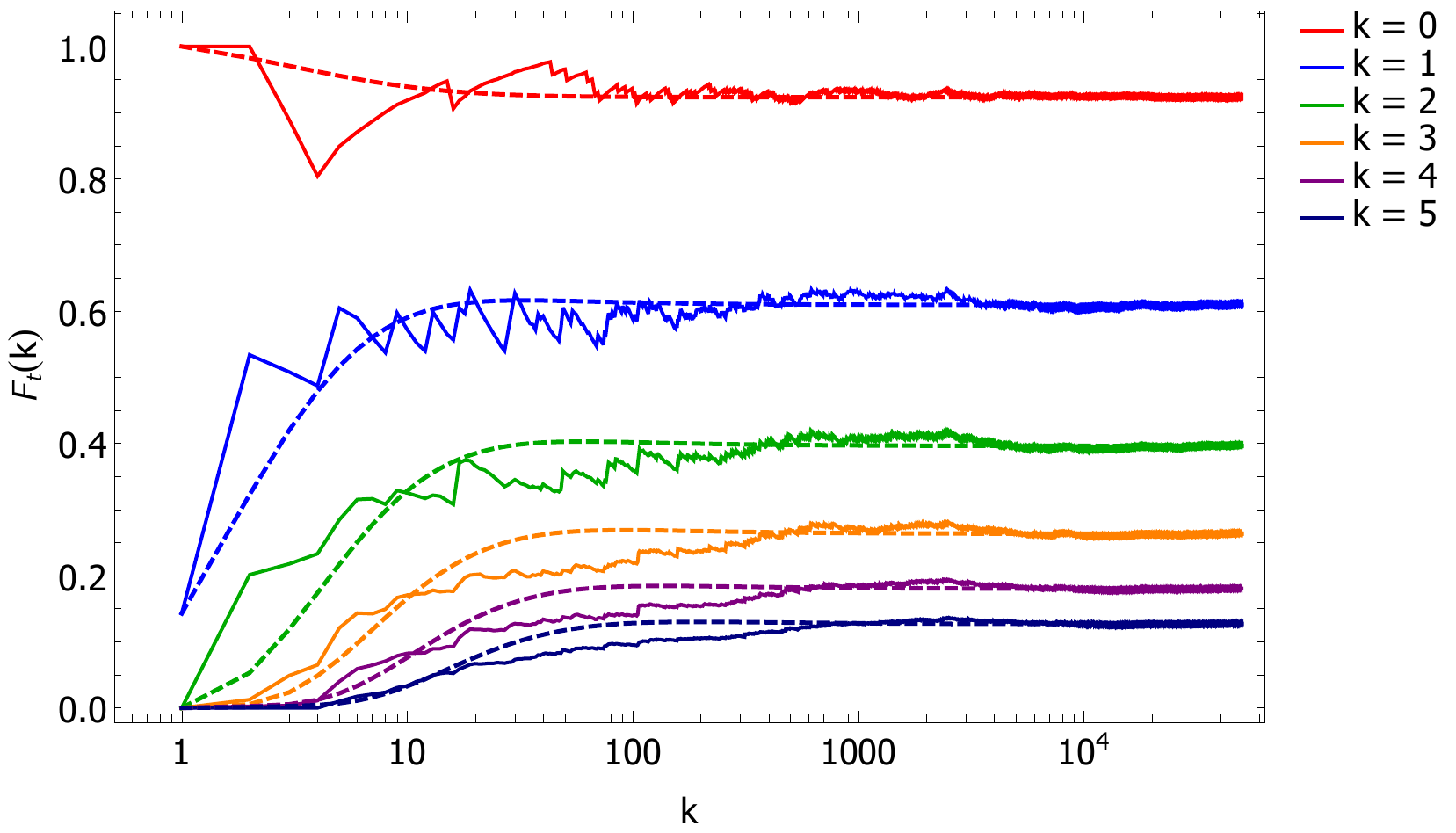}}
    
    \subfloat[][]{\includegraphics[width=200pt,height=120pt]{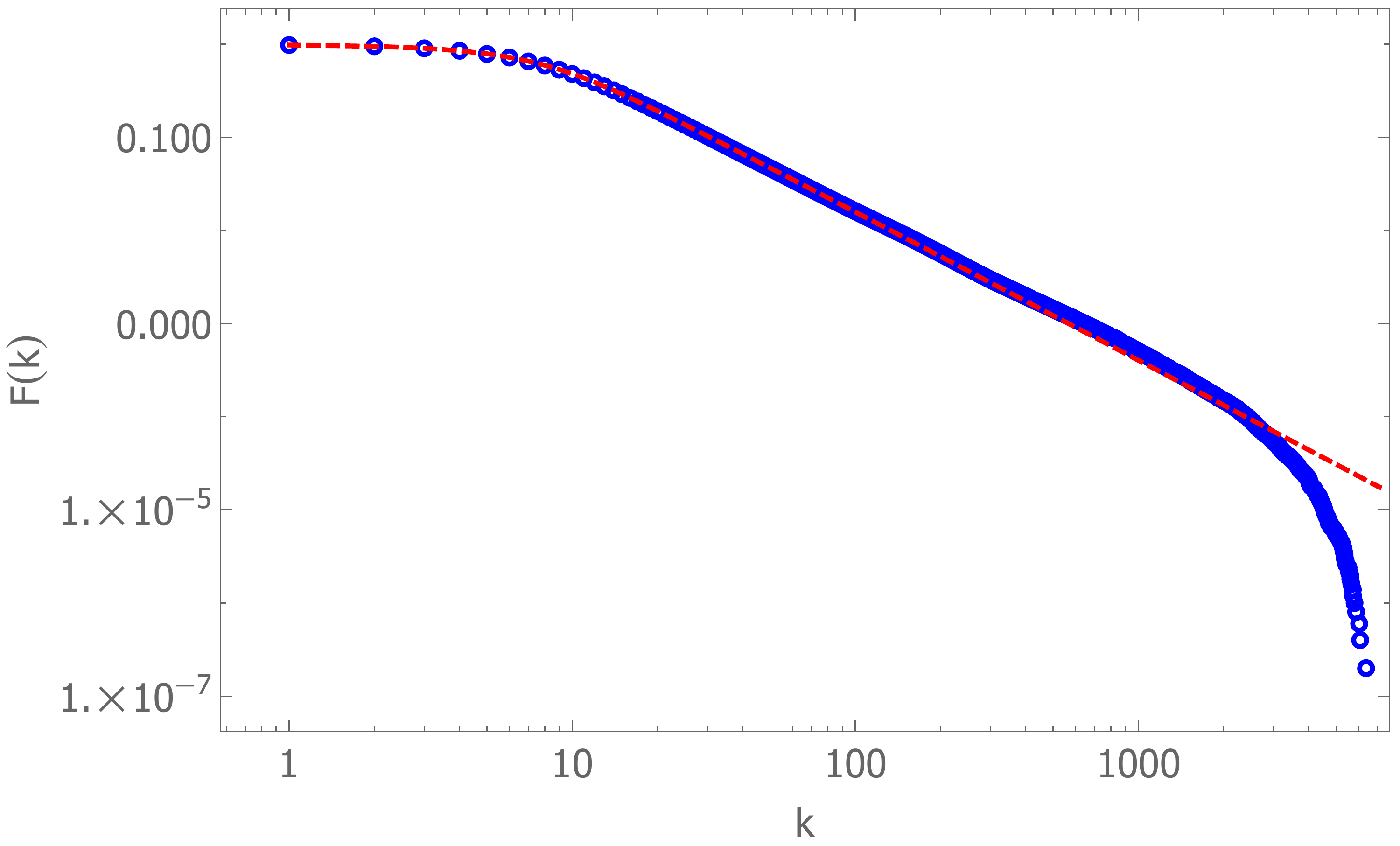}}~
    \subfloat[][]{\includegraphics[width=200pt,height=120pt]{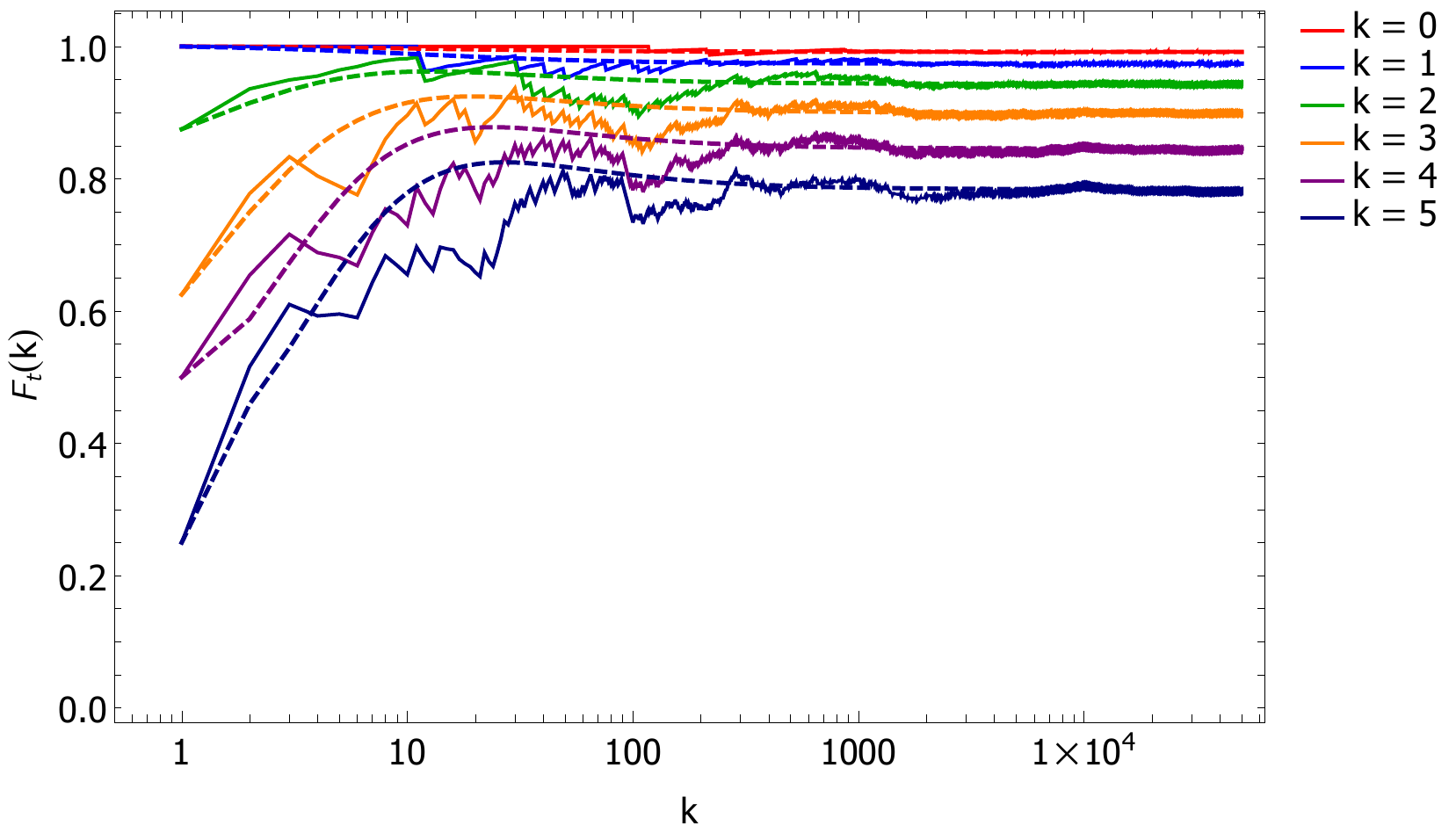}}
    \caption{Complementary cumulative in-degree distribution for three growing networks. The dots represent an average
of 150 simulation runs and the solid lines indicate the theoretical predictions based on \cref{eq3.11,eq3.12,eq3.13}. Networks are generated with (a)-(b) $p=0$, $q=0.4$, and $A$ follows a geometric distribution with $\theta=0.3$; (c)-(d) $p=0.5$, $q=0.8$, and $A$ follow a log-series distribution with $\theta=0.6$; (e)-(f) $p=1$, $q=0.6$, and $A$ follows a negative binomial distribution with $r=15.2$ and $\theta=0.4$.}
    \label{fig1}
\end{figure}

\cref{fig1} shows the behavior of a simulated network with $10^3$ nodes and the theoretical predictions based on \cref{theorem3.2,theorem3.3}. The left plot illustrates the convergence and the right plot the dynamics of $F_t$ for nodes with in-degree $k=0,1,\ldots,5$.

%section3
\section{Results}\label{section4}
This section illustrates an application of~\cref{theorem3.1,theorem3.2}. In particular, it examines the effect of the distribution of the non-constant growth and mechanism M3 on the in-degree distributions of networks.\\

\cref{theorem3.2} shows that the theoretical CCDF depends in general on three parameters: the distribution of new edges, the proportion of preferential attachment ($p$), and the proportion of reciprocity ($q$). In general, determining the best-fit parameters for empirical data is a knotty task. The following procedure summarizes the key steps for estimating the parameters.

\begin{enumerate}
\item Estimate the parameter $k_{\mathrm{sat}}$ (based on the approximation in~\cite{Clauset2009}) and split the distribution into two parts (head and tail).
\item Apply maximum likelihood estimation (MLE) to fit the head of the distribution.
\item Calculate the goodness-of-fit between the empirical data and the network model. A variant of the Kolmogorov-Smirnov statistic (KS statistic), which measures the maximum distance between the CCDF of the in-degree of the empirical network and the fitted model, can be applied. In particular,
\[\mathrm{KS}=\max _{x\in\mathrm{dom}(S)} \mid S(x)-F(x)\mid\]
where $S(x)$ represents CCDF of the empirical network, $F(x)$ the CCDF of the network model, and $\mathrm{dom}(S)$ the support of $S$.
\end{enumerate}

In the following subsections, we apply our approach to two empirical networks. For the first network timestamps for the creation of new edges are not available (we do not know when the edges was created), so we focus on finding the best fit for the final in-degree distribution. For the second network, since we are able to evaluate the evolution of the CCDF when fitting the model to the data.

% would prefer if we stick with F(c) for the distribution without having  to introduce the CDF

\subsection{Free On-line Dictionary of Computing (FOLDOC)}
Consider the dataset of the Free On-line Dictionary of Computing (FOLDOC). Nodes represent dictionary terms (entries) and a directed edge from term $u$ to term $v$ denotes that $v$ is used in the definition of $u$~\cite{Batagelj2002}. The resulting network is directed and consist of $13.3$ thousand nodes and $125.2$ thousand edges. Self-loops and multi-edges are neglected.\\

\begin{figure}[hbtp!]
    \centering
    \subfloat[][]{\includegraphics[width=200pt,height=120pt]{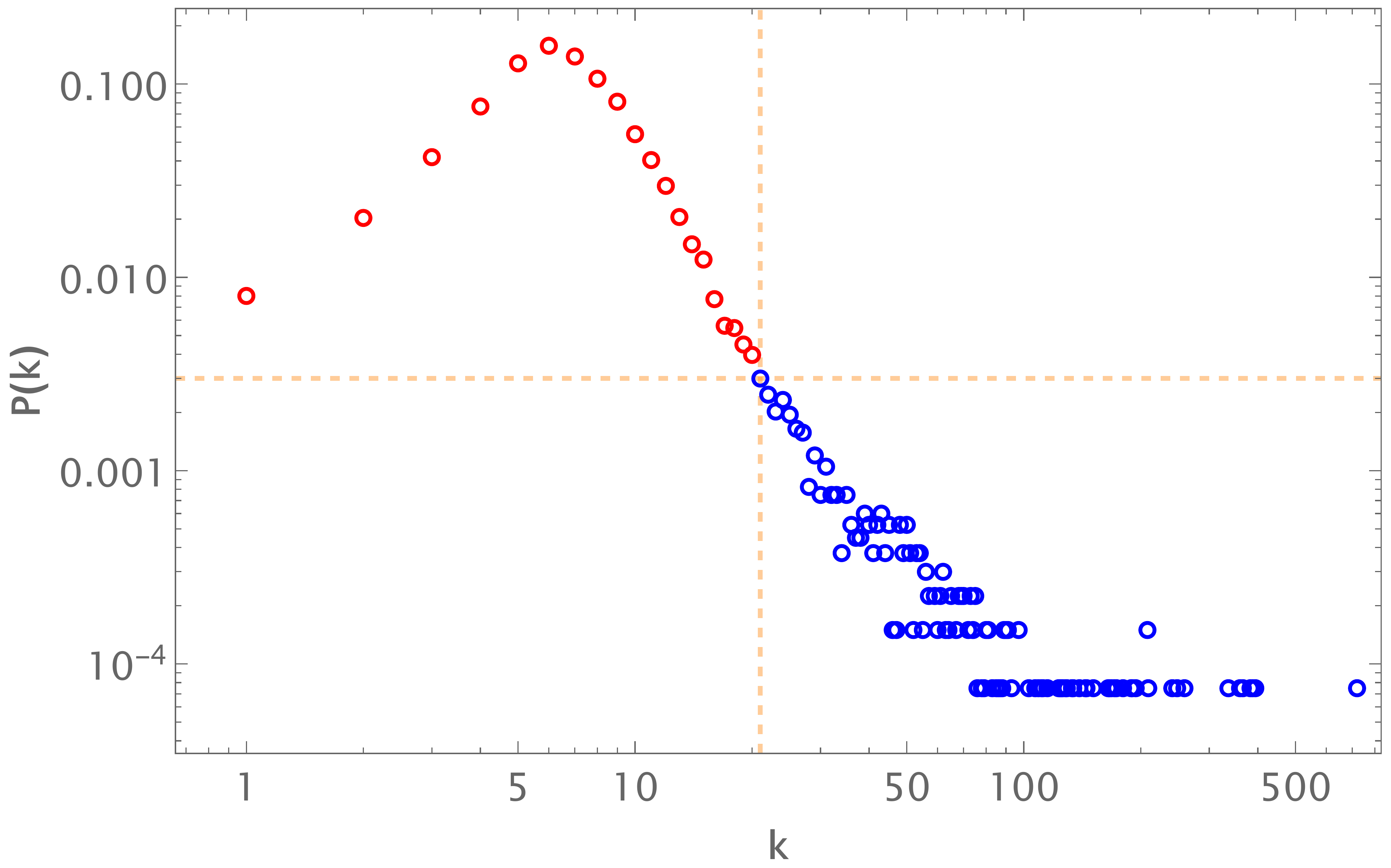}\label{fig2a}}~
    \subfloat[][]{\includegraphics[width=200pt,height=120pt]{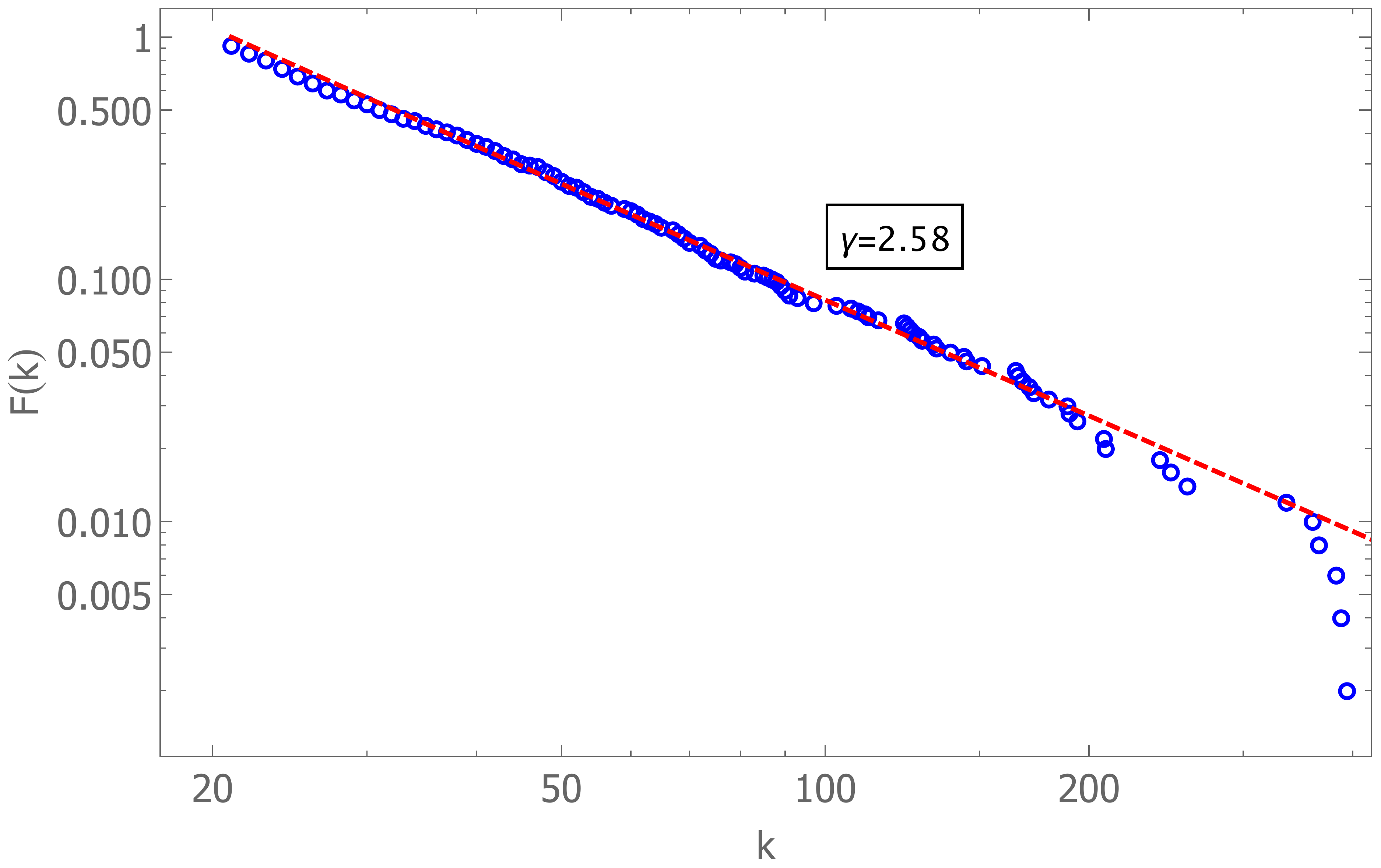}\label{fig2b}}
    \caption{(a) In-degree distribution of the FOLDOC network. The head and the tail of the distribution are depicted by the red and blue circles, respectively. (b) Tail of the CCDF of the in-degree.}
    \label{fig2}
\end{figure}

\Cref{fig2a} shows that the in-degree distribution of the network is a mixture of two different distributions. To differentiate between them, we split the in-degree data into two sets. The first set corresponds to the number of nodes with in-degree less than $k_{\mathrm{sat}}=21$. The second set corresponds to the number of nodes with an in-degree greater or equal than $k_{\mathrm{sat}}$. The resulting distributions correspond to the head and tail of the in-degree distribution of the FOLDOC network. The tail follows a power-law with $\gamma=2.58$~(\cref{fig2b}). 

\begin{figure}[hbt!]
    \centering
    \subfloat[][]{\includegraphics[width=220pt]{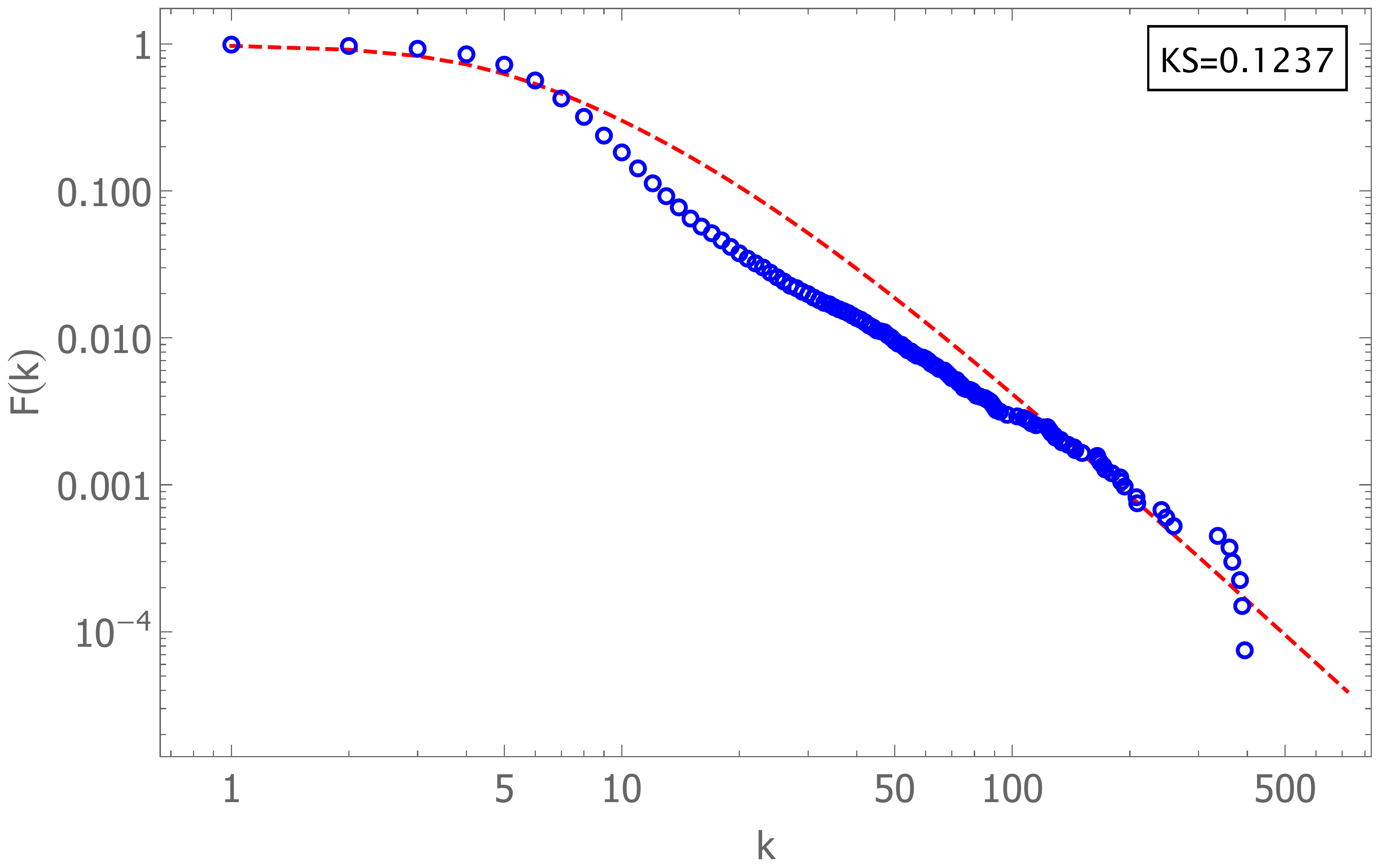}}~
    \subfloat[][]{\includegraphics[width=220pt]{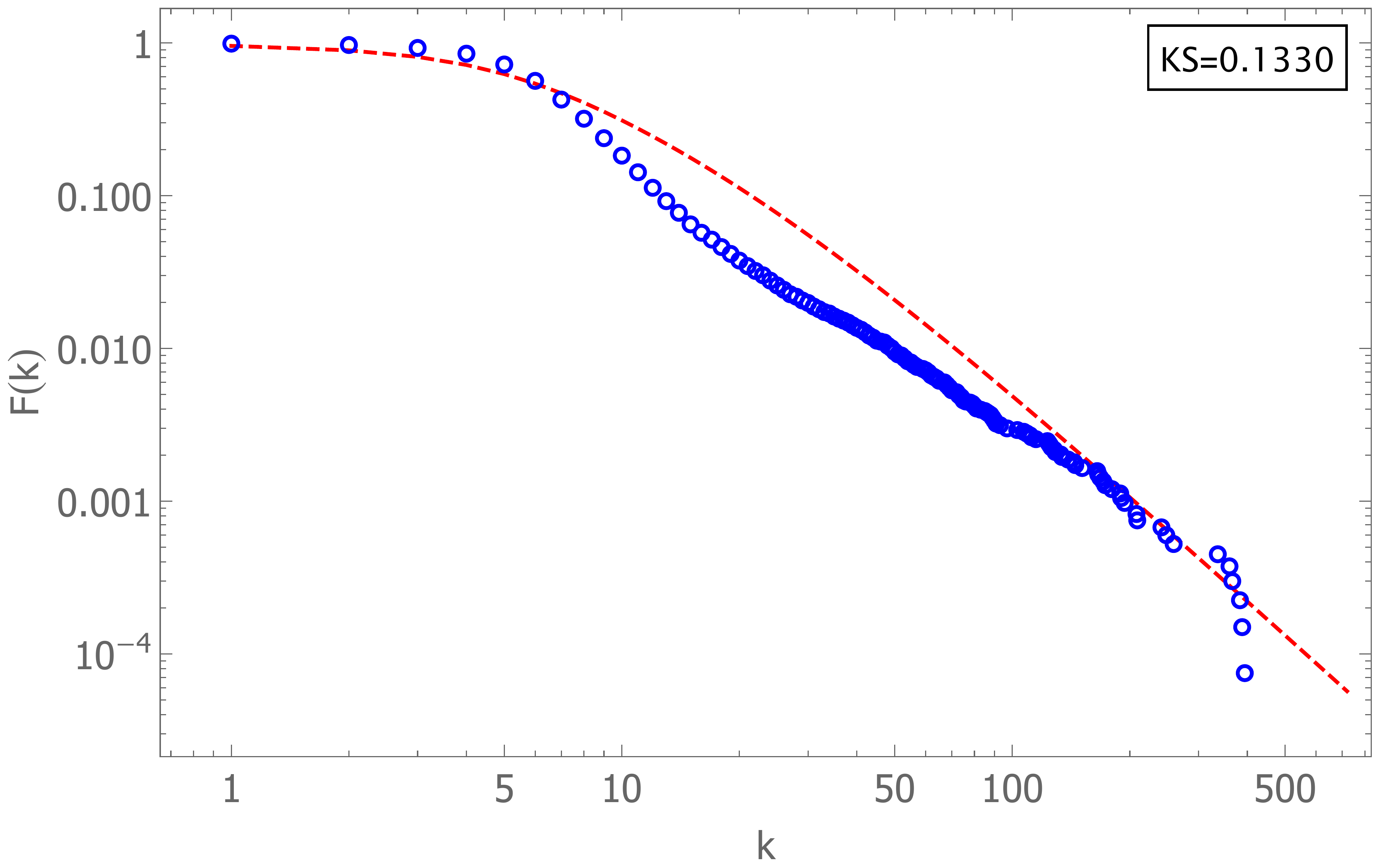}}
    
    \subfloat[][]{\includegraphics[width=220pt]{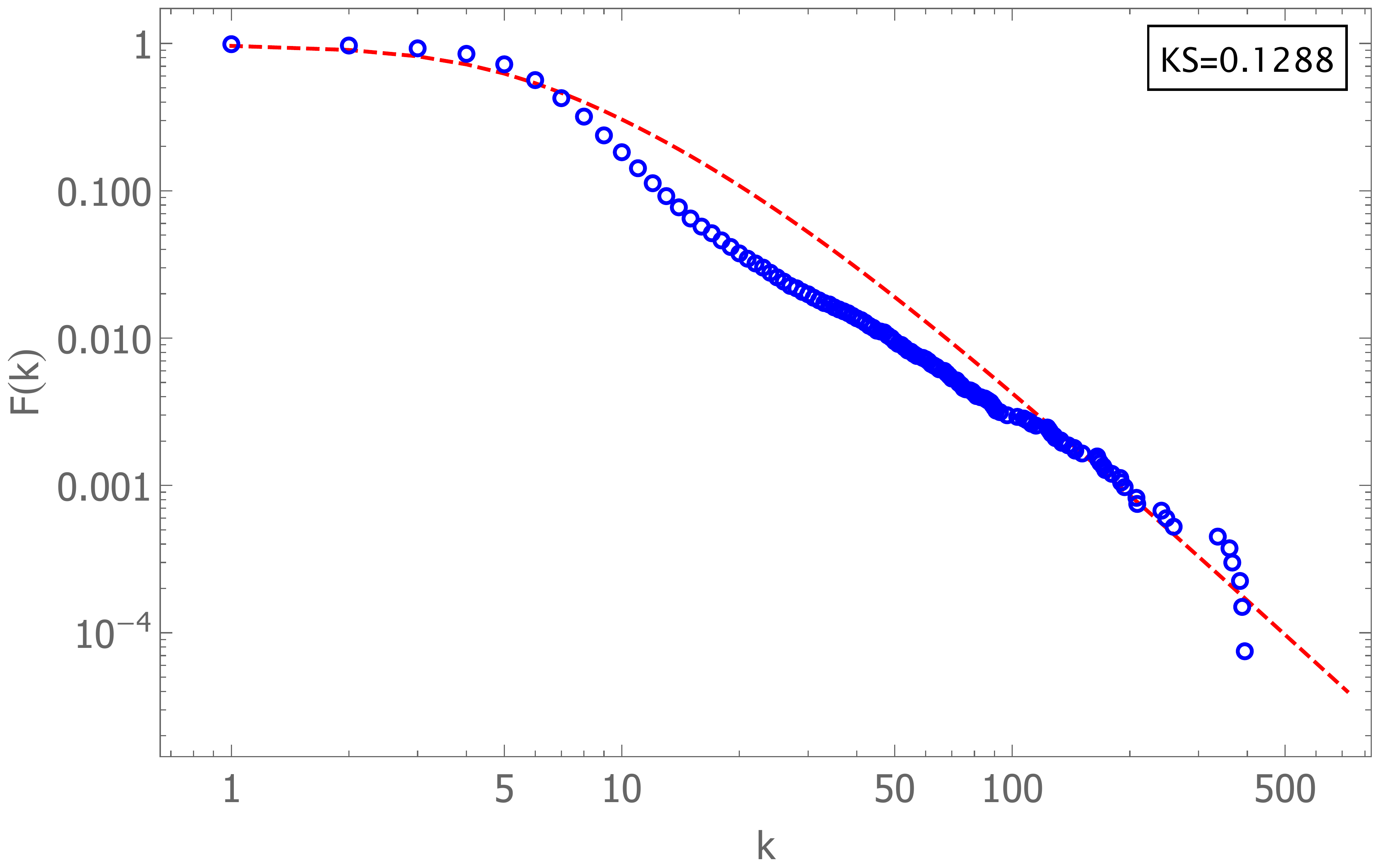}}
\caption{Goodness-of-fit between the data and the fitted model. (a)~The random variable $A$ follows a binomial distribution with $p=0.6$; (b)~$A$ follows a negative binomial with $p=0.63$; (c)~$A$ follows a Poisson distribution with $p=0.6$.}
\label{fig3}
\end{figure}

Next, to characterize the distribution of the head, we apply a maximum likelihood estimator to fit the data to three discrete probability distributions: Binomial, Negative Binomial, and Poisson. \cref{table1} summarizes the estimated parameters for each distribution.
\begin{table}[!hbt]
\centering
\begin{tabular}{|l|l|l|}
\hline
\multicolumn{1}{|c|}{Distribution} & \multicolumn{2}{c|}{Parameters} \\ \hline
Binomial                           & $r=20$        & $\theta=0.37$   \\ \hline
Negative Binomial                  & $r=16.92$    & $\theta=0.69$   \\ \hline
Poisson                            & \multicolumn{2}{c|}{$r=7.43$}   \\ \hline
\end{tabular}
\caption{Estimated parameters by MLE.}
\label{table1}
\end{table}
Assuming that the proportion of reciprocal edges does not vary over time, the proportion of reciprocity is estimated as $q=0.47$. Finally, using the KS statistic, we tune the parameter $p$. \cref{fig3} illustrates the final fitting results.

The estimated values have the best fit when $p = 0.6$ and the random variable $A$ follows a binomial distribution with parameters $(20,0.37)$. For each case in \cref{fig3}, the proportion of preferential attachment is greater than or equal to $0.6$. This result suggests that as new terms are defined based on existing terms, approximately $60\%$ of the definitions are composed of popular terms in the dictionary.

\subsection{Stack Overflow A2Q Network}
The Stack Overflow A2Q Network captures user interactions of questions and answers on Stack Overflow~\cite{Paranjape2017}. The original dataset consists of $2.46$ million questions and $17.82$ million answers. Let a node represent a user who posted or answered at least one question on Stack Overflow. An edge represents an answer from a user to a question by another user. As a seed network, consider the network $ G_0 = (V_0, E_0) $, composed of nodes representing users who published a question or an answer on August 1, 2008 between 0:00 am to 5:00 pm. This network $G_0$ has $33$ nodes and $35$ edges. There is a directed edge $(u,v)\in E_0$ from user $u$ to user $v$ if and only if user $u$ provided an answer to a question posted by user $v$, with both actions occurring within the 17-hour period.\\

The sequence $(G_t)_T$ is built by sorting in ascending order the time-stamps associated with users who are not present in $V_0$. The network grows by the continuous addition of nodes and edges with time-stamps up to March 6, 2016. In particular, from the seed network $G_0$, at each time step $t > 0$, the network $G_t = (V_t, E_t)$ is constructed according to the following steps.

\begin{enumerate}
\item The set of nodes $V_t$ contains all nodes in $V_{t-1}$ plus a new node, $u_t$, which represents the user that answers at time $t$ a question formulated by a user in $V_{t-1}$; and
\item The set of edges $E_t$ contains all edges in $E_{t-1}$ plus a new edge, $(u_t, v)$, for each user $v\in V_t$ for each question of $v$ answered by $u_t$.
\end{enumerate}

The above steps result in a sequence $(G_t)_T$ of networks, with approximately $7.66$ thousand nodes and $8.00$ thousand edges. The sequence of networks represents the growth of the A2Q network from 2008 to 2016.\\

\begin{figure}[!hbtp]
    \centering
    \subfloat[][]{\includegraphics[width=200pt,height=120pt]{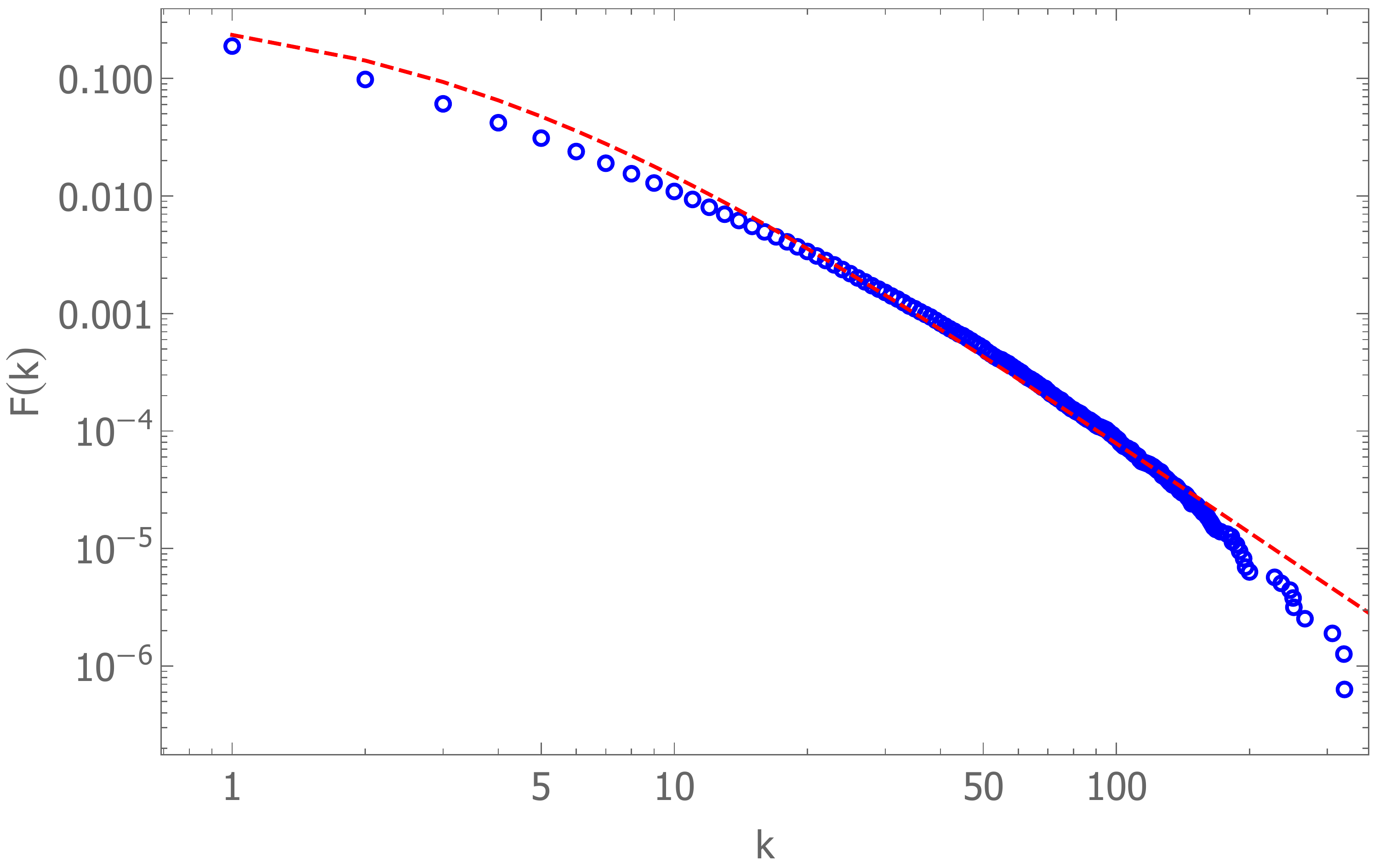}\label{fig4a}}~
    \subfloat[][]{\includegraphics[width=200pt,height=120pt]{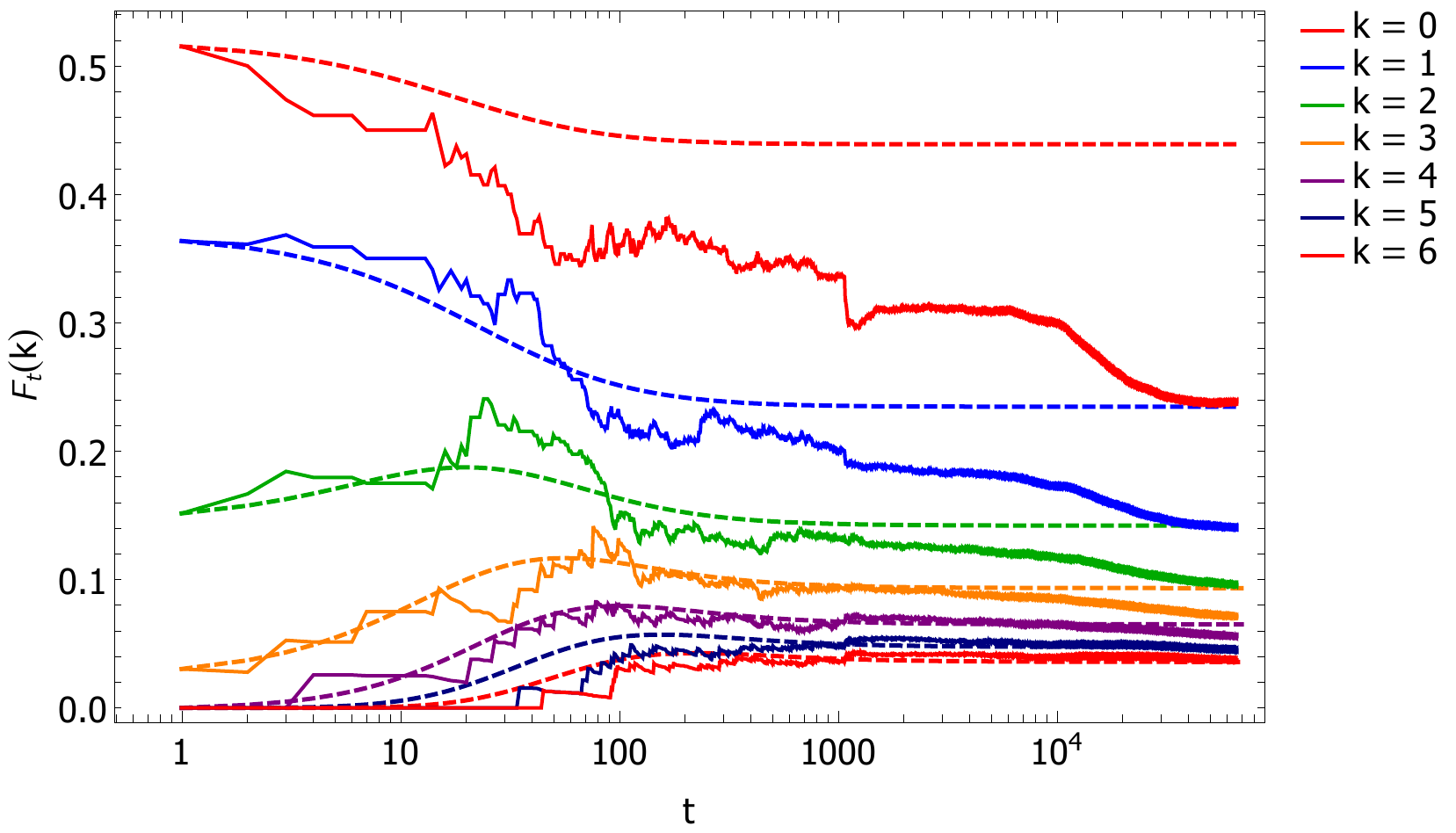}\label{fig4b}}
    
\caption{(a) Complementary cumulative degree distribution for a mix attachment network with parameters $p=0.39$ and $q=0.01$. The dots represent the empirical distribution and the dashed line indicates the theoretical prediction based on \cref{theorem3.2}. (b) Evolution of the in-degree distribution for nodes with in-degrees $k = 0,1,\ldots, 6$ based on \cref{theorem3.3}.}
\label{fig4}
\end{figure}

Next, based on the number of links generated by each new node, we fit the data to a Zeta distribution with parameter $\theta=2.29$ (the parameter is estimated using MLE). \cref{fig4} shows the theoretical predictions based on \cref{theorem3.2,theorem3.3}.\\

\cref{fig4b} indicates that while a good fit for the CCDF is attained, there are significant deviations between the empirical and theoretical dynamics of the in-degree. These deviations can be explained by the assumption that the extent to which preferential attachment dominates the attachment process, as well reciprocity the response mechanism, remain constant. A more realistic scenario requires us to extend the model and its analytical results to allow for $p$ and $q$ to vary over time.%section4
\section{Conclusions}\label{section5}

Our work presents a network model with a mixed attachment mechanisms with varying number of new edges and a response mechanism that allows existing nodes to respond to such attachments by establishing reciprocal edges. We capture the effects of the number of new edges and reciprocity as the network grows by the continuous addition of new nodes. Moreover, we characterize the distribution of the in-degree of new nodes. Based on these characterizations, we derive analytical expressions for the dynamics and limit behavior of the in-degree distribution of the resulting network.\\

These results enhance our understanding of the formation of extended exponential and extended power-law networks. In particular, they enable us to take into account the effects of the process of reciprocity on the head of in-degree distributions. Allowing for time-varying mixed attachment and response mechanisms remains an interesting direction for future research.

%section
\section*{Acknowledgments}
This research was supported in part by the Colombian ``Fondo de Ciencia, Tecnología e Innovación del Sistema General de Regalías FCTeI-SGR'' of the Cauca Department throughout the project ``Fortalecimiento de las Capacidades de las EBT-TIC del Cauca para Competir en un Mercado Global - Cluster CreaTic'' under grant no. TH 2017-01.
\bibliographystyle{ieeetr}
\bibliography{references}

\newpage
\section*{Appendix: Proof of \cref{theorem3.1}}
\begin{proof}
\begin{enumerate}

\item 
	Since $A$ follows a Poisson distribution with parameter $r$, we have
	\begin{equation*}
	\mathbb{P}(A=i)=\frac{r^i}{i!}e^{-r}.
	\end{equation*}
	By using \cref{eq3.1}, we get
	\begin{align*}
	\mathbb{B}(k)&=\sum_{i=k}^{\infty}\binom{i}{k}q^k(1-q)^{i-k}~\mathbb{P}(A=i)\\
	&=\frac{q^k}{k!}\sum_{i=k}^{\infty}\left(\frac{i!}{(i-k)!}(1-q)^{i-k}\right)\left(\frac{r^i}{i!}e^{-r}\right)\\
	&=\frac{(r q)^k}{k!}e^{-r}\sum_{i=k}^{\infty}\frac{(r(1-q))^{i-k}}{(i-k)!},
	\end{align*}
	Note that the last sum in the above equation is the Taylor series for $e^{r(1-q)}$. So, we have
	\begin{align*}
	\mathbb{B}(k)=\frac{(qr)^k}{k!}e^{-qr}.
	\end{align*}
\item 
	Since $A$ follows a binomial  distribution with parameters $r$ and $\theta$, we have
	\begin{equation*}
	\mathbb{P}(A=i)=\binom{r}{i}\theta^i(1-\theta)^{r-i}.
	\end{equation*}
	By using \cref{eq3.1}, we get
	\begin{align*}
	\mathbb{B}(k)&=\sum_{i=k}^{\infty}\binom{i}{k}q^k(1-q)^{i-k}~\mathbb{P}(A=i)\\
	&=\sum_{i=k}^{\infty}\left(\frac{i!}{(i-k)!~k!}q^k(1-q)^{i-k}\right)\left(\frac{r!}{(r-i)!~i!}\theta^i(1-\theta)^{r-i}\right)\\
	&=\frac{r!}{k!}q^k\sum_{i=k}^{\infty}\frac{1}{(i-k)!~(r-i)!}\theta^i(1-q)^{i-k}(1-\theta)^{r-i}.
	\end{align*}
	Letting $j=i-k$, we have
	\begin{align*}
	\mathbb{B}(k)&=\binom{r}{k}(q\theta)^k(1-\theta)^{r-k}\sum_{j=0}^{\infty}\binom{r-k}{j}\left(\frac{\theta(1-q)}{1-\theta}\right)^j,
	\end{align*}
	and assuming that $\left|\frac{\theta(1-q)}{1-\theta}\right|<1$, the last sum is the binomial series of $\left(1+\frac{\theta(1-q)}{1-p}\right)^{r-k}$. So, we have
	\begin{align*}
	\mathbb{B}(k)=\binom{r}{k}(q\theta)^k(1-q\theta)^{r-k}.
	\end{align*}
\item
	Since $A$ follows a geometric distribution with parameter $\theta$, we have
	\begin{equation*}
	\mathbb{P}(A=i)=\theta(1-\theta)^{i}.
	\end{equation*}
	By using \cref{eq3.1}, we get
	\begin{align*}
	\mathbb{B}(k)&=\sum_{i=k}^{\infty}\binom{i}{k}q^k(1-q)^{i-k}~\mathbb{P}(A=i)\\
	&=\sum_{i=k}^{\infty}\left(\frac{i!}{(i-k)!~k!}q^k(1-q)^{i-k}\right)\left(\theta(1-\theta)^{i}\right)\\
	&=\theta\left(\frac{q}{1-q}\right)^k\sum_{i=k}^{\infty}\frac{i!}{(i-k)!~k!}\left((1-q)(1-\theta)\right)^i.
	\end{align*}
	Letting $j=i-k$,
	\begin{align*}
	\mathbb{B}(k)&=\theta\left(q(1-\theta)\right)^k \sum_{j=0}^{\infty}\binom{j+k}{k}\left((1-q)(1-\theta)\right)^{j},
	\end{align*}
	and assuming that $q+\theta(1-q)\neq 0$, the last sum is the binomial series of $\left(1-(1-q)(1-\theta)\right)^{-(k+1)}$. So, we have
	\begin{align*}
	\mathbb{B}(k)=\frac{\theta}{q+\theta(1-q)}\left(\frac{q(1-\theta)}{q+\theta(1-q)}\right)^k.
	\end{align*}
\item 
	Since $A$ follows a negative binomial distribution with parameters $r$ and $\theta$, we have
	\begin{equation*}
	\mathbb{P}(A=i)=\binom{r+i-1}{r-1}\theta^i(1-\theta)^{r}.
	\end{equation*}
	By using \cref{eq3.1}, we get
	\begin{align*}
	\mathbb{B}(k)&=\sum_{i=k}^{\infty}\binom{i}{k}q^k(1-q)^{i-k}~\mathbb{P}(A=i)\\
	&=\sum_{i=k}^{\infty}\left(\frac{i!}{(i-k)!~k!}q^k(1-q)^{i-k}\right)\left(\frac{(r+i-1)!}{(r-1)!~i!}\theta^i(1-\theta)^{r}\right).
	\end{align*}
	Letting $j=i-k$, we have
	\begin{align*}
	\mathbb{B}(k)&= (q\theta)^k(1-\theta)^r \binom{k+r-1}{k}\sum_{j=0}^{\infty}\binom{j+k+r-1}{j}\left(\theta(1-q)\right)^j,
	\end{align*}
	and assuming that $q+\theta(1-q)\neq 0$, the last sum is the binomial series of $\left(1-\theta(1-q)\right)^{-(k+r)}$. So, we have
	\begin{align*}
	\mathbb{B}(k)=\binom{k+r-1}{k}\left(\frac{q\theta}{1-\theta(1-q)}\right)^k\left(\frac{1-\theta}{1-\theta(1-q)}\right)^r.
	\end{align*}
\item
	Since $A$ follows a zeta distribution with parameter $r$, we have
	\begin{equation*}
	\mathbb{P}(A=i)=\frac{i^{-(r+1)}}{\zeta(r+1)}.
	\end{equation*}
	By using \cref{eq3.1}, we get
	\begin{align*}
	\mathbb{B}(k)&=\sum_{i=k}^{\infty}\binom{i}{k}q^k(1-q)^{i-k}~\mathbb{P}(A=i)\\
	&=\sum_{i=k}^{\infty}\left(\frac{i!}{(i-k)!~k!}q^k(1-q)^{i-k}\right)\left(\frac{i^{-(r+1)}}{\zeta(r+1)}\right).
	\end{align*}
	For $k=0$,
	\begin{align*}
	\mathbb{B}(0) &= \frac{1}{\zeta(r+1)}\sum_{i=0}^{\infty} \frac{(1-q)^i}{i^{r+1}},
	\end{align*}
	and assuming that $r\geq 1$ and $|1-q|<1$, the last sum is the series representation of the poly-logarithm function. So, we have
	\begin{align*}
	\mathbb{B}(0)=\frac{1}{\zeta(r+1)}\mathrm{Li}_{r+1}(1-q).
	\end{align*}
	For $k\geq 1$,
	\begin{align*}
	\mathbb{B}(k)&=\sum_{i=0}^{\infty}\left(\frac{i!}{(i-k)!~k!}q^k(1-q)^{i-k}\right)\left(\frac{i^{-(r+1)}}{\zeta(r+1)}\right)\\
	&= \frac{1}{\zeta(r+1)}\left(\frac{q}{1-q}\right)^k\sum_{i=k}^{\infty}\frac{(1-q)^i}{i^{r+1}}\left(\frac{(i-k+1)(i-k+2)\cdots(i-2)(i-1)i}{k!}\right),
	\end{align*}
	as before assuming $r\geq 1$ and $|1-q|<1$, we get
	\begin{align*}
	\mathbb{B}(k)=\frac{1}{\zeta(r+1)k!}\left(\frac{q}{1-q}\right)^k\sum_{i=1}^{k}e_{i-1}(1,\ldots,k)\mathrm{Li}_{r+i-k}(1-q),
	\end{align*}
	where $e_i(1,\ldots,k)$ denotes the elementary symmetric polynomials in $k$ variables.
\item
	Since $A$ follows a log series distribution with parameter $\theta$, we have
	\begin{equation*}
	\mathbb{P}(A=i)=-\frac{\theta^i}{i \log(1-\theta)}.
	\end{equation*}
	By using \cref{eq3.1}, we get
	\begin{align*}
	\mathbb{B}(k)&=\sum_{i=k}^{\infty}\binom{i}{k}q^k(1-q)^{i-k}~\mathbb{P}(A=i)\\
	&=\sum_{i=k}^{\infty}\left(\frac{i!}{(i-k)!~k!}q^k(1-q)^{i-k}\right)\left(-\frac{\theta^i}{i \log(1-\theta)}\right).
	\end{align*}
	Since $\mathbb{P}(A=0)=0$, for $k=0$
	\begin{align*}
	\mathbb{B}(0)&= \sum_{i=1}^{\infty} (-1)^{i+1}\frac{\left(\theta(q-1)\right)^i}{i \log(1-\theta)}.
	\end{align*}
	Assuming that $\theta\neq 1$, the last sum is the Newton–Mercator series. So, we have
	\begin{align*}
	\mathbb{B}(0)=\frac{\log(1+\theta(q-1))}{\log(1-\theta)}.
	\end{align*}
	For $k\geq 1$, and letting $j=i-k$
	\begin{align*}
	\mathbb{B}(k)&=\sum_{j=0}^{\infty}\left(\frac{(j+k)!}{j!~k!}q^k(1-q)^{j}\right)\left(-\frac{\theta^{j+k}}{(j+k) \log(1-\theta)}\right)\\
	&= -\frac{(q\theta)^k}{k\log(1-\theta)}\sum_{j=0}^{\infty}\binom{j+k-1}{j}\left(\theta(1-q)\right)^j,
	\end{align*}
	as before assuming $\theta\neq 1$, the last sum is a binomial series. Therefore
	\begin{align*}
	\mathbb{B}(k)=-\frac{1}{k\log(1-\theta)}\left(\frac{q\theta}{1-\theta(1-q)}\right)^{k}
	\end{align*}
\end{enumerate}
\end{proof}

\end{document}